\documentclass[]{JHEP3} 

\usepackage{epsfig,multicol,bbm,cite}

\def\be{\begin{equation}}
\def\ee{\end{equation}}
\def\bea{\begin{eqnarray}}
\def\eea{\end{eqnarray}}

\def\re{\mathrm{Re}}
\def\Tr{\mathrm{Tr}}

\newcommand{\sect}[1]{Sect.~\ref{#1}}
\newcommand{\fig}[1]{Fig.~\ref{#1}}
\newcommand{\tab}[1]{Table~\ref{#1}}

\title{QCD with one compact spatial dimension}

\author{Thomas DeGrand and Roland Hoffmann\\
	Department of Physics, University of Colorado,Boulder, CO 80309 USA\\
	E-mail: \email{degrand@pizero.colorado.edu}, \email{hoffmann@pizero.colorado.edu}}
\received{February 20, 2001} 		
\revised{May 1, 2001}
\accepted{November 27, 2001}		

\abstract{The realization of global symmetries can depend on the geometry
of the underlying space. In particular, compactification can lead
to spontaneous breaking of such symmetries.
Four--dimensional QCD with fundamental representation fermions
embedded in a space with one compact spatial dimension has a critical length, at which the
theory undergoes a phase transition and develops a ground state that is
no longer charge conjugation invariant. We show this behavior with 
simulations of three color, four flavor
QCD. We use
unrooted staggered fermion at two values of the lattice spacing and
several quark masses.
We discuss the dependence of the transition on the dynamical fermion mass as well
as its connection to the finite temperature and chiral phase transitions.}


\begin{document} 

\section{Introduction}

\subsection{Motivation}

When field theories are defined on spaces with one or more compact directions,
the sizes of the compact dimensions can influence the existence and location
of phase transitions. The most familiar example of this
behavior is the use of a compact temporal dimension to study a field theory at
finite temperature. In this paper we describe such a situation, which occurs in ordinary
QCD (with dynamical fermions): a sufficiently small compact spatial dimension
induces a transition into a phase in which charge conjugation is
spontaneously broken. The existence of this new phase, and of the length scale which 
characterizes the transition, depend on the choice of periodic fermionic boundary conditions
in the compact dimension(s).

The effect we describe is similar to the deconfinement transition in QCD with dynamical
fermions. The action of
$SU(N)$ pure gauge theories is invariant under gauge transformations which
are periodic up to multiplication by the center of the gauge group, $Z(N)$ for $SU(N)$.
These theories have an order parameter, the Polyakov loop or Wilson line, which is 
sensitive to such transformations. The low temperature phase of the theory is one
in which the vacuum expectation value of the order parameter vanishes. In the high temperature
phase, the order parameter spontaneously magnetizes along one of the directions 
of the center of the gauge group, signaling deconfinement.
Now add dynamical fermions.
The Euclidean path integral for fermions has a probability interpretation when the
temporal boundary conditions of the fermions are chosen to be anti--periodic. The fermion
action (for fundamental representation fermions) explicitly breaks the $Z(N)$ symmetry,
favoring the vacuum in which
the Polyakov loop takes a positive value.

For the physical case of three colors, the deconfinement transition is first order
for the pure gauge theory and is believed to disappear as the fermion mass is reduced
from infinity, as only one vacuum configuration is energetically favored.
This is all well known: for an early review, see
 \cite{Svetitsky:1985ye}.

However, if the fermions obey periodic boundary conditions, the vacuum at the positive
center element is disfavored. If the number of colors is odd, there
are two degenerate minima in the free energy, corresponding to complex-conjugate
values of the order parameter. One of these is selected as the vacuum
and thus charge conjugation is spontaneously broken. The radius of
the compact dimension at which the transition from $C$-conserving to $C$-broken
occurs is a physical property of QCD.

The earliest discussion of this new phase that we are aware of is by van Baal
\cite{vanBaal:1988va,vanBaal:2000zc}. He was interested in QCD with all three
spatial dimensions taken to be small. In this case there are $2^3$
degenerate vacua, two per small dimension.
While he suggested that numerical simulations of QCD with compact dimensions
might be interesting, we are aware of no follow-up studies.

Our interest in this problem was initiated from a quite different source:
a discussion of the validity of  the ``orientifold
QCD'' programme of Armoni, Shifman, and Veneziano
\cite{Armoni:2003gp,Armoni:2003fb,Armoni:2003yv}. These authors proposed a large-$N_c$
 equivalence between
$\mathcal{N}\!=\!1$ super Yang-Mills and QCD with
a single fermion flavor in the symmetric or antisymmetric tensor representation.
The latter theory is equivalent to one-flavor QCD for $N_c=3$.
The authors used this connection to make a successful prediction of the value of the condensate of
one-flavor QCD \cite{Armoni:2003yv,DeGrand:2006uy}. 

This summer, \"Unsal  and Yaffe \cite{Unsal:2006pj} pointed out that a necessary
condition for the non--perturbative equivalence of different theories is that global
symmetries are realized
in the same manner in both.
They show that if the theories are considered on $\mathbb{R}^3\times\mathcal{S}^1$ and the radius of
$\mathcal{S}^1$ is chosen sufficiently small, QCD with tensor representation fermions will
spontaneously break charge conjugation symmetry, while the supersymmetric theory does not.
This work is a specific application of their earlier, more general studies
\cite{Kovtun:2003hr,Kovtun:2004bz,Kovtun:2005kh}.

Their discussion is very general: it depends only on the gauge group and on the representation
of the fermions. We decided to look for the $C$-violating phase in numerical simulations,
and to do this for a choice of parameters which was computationally the most inexpensive.
Accordingly, we performed simulations with three colors and four flavors of fundamental
representation dynamical fermions. We concentrate on zero-temperature simulations with
one compact dimension (this is done in the simulation by taking the compact dimension much
smaller than the other ones) but also briefly consider other situations: finite temperature
plus one or more compact spatial dimensions. The transition is strong and easy to observe.
We have not attempted to determine its order.

In the remainder of this section we outline theoretical expectations for $C$-odd phases, and then
proceed to simulations and their results.

\subsection{Weak and strong coupling}\label{sec:coup}

Asymptotic freedom ensures that a perturbative calculation of the effective potential
for Polyakov lines becomes reliable if their length $L$ is short enough. In this case
the coupling at the scale $1/L$ is small and higher order corrections become negligible.
For $n_f$ flavors of massless
Dirac fermions in the fundamental representation of $SU(N)$ the effective 
potential for the Polyakov loop $P$ is given by \cite{Unsal:2006pj}
\be
V_{\rm eff}(P)=-\frac1{LV}\log Z[P]=-\frac1{LV}\log\left[
\frac{\det^{2n_f}(-D^2_{\rm fund})}
     {\det(-D^2_{\rm adj})}
\right]\;,\label{Veff}
\ee
where $D^2$ is the covariant Laplacian for a constant background field in a given representation
and $V$ denotes the volume of the non--compact dimensions.
An evaluation of (\ref{Veff}) proceeds
exactly as in Ref.~\cite{Unsal:2006pj};
the (positive) contribution to $V_{\rm eff}$ from the bosonic degrees of freedom
(in a gauge where the Polyakov loop in the compact direction is
$\mathrm{diag}(e^{iv_1},\ldots,e^{iv_N})$) is minimized
when $v_j\!=\!v$ for all $j$ and one obtains for periodic/anti--periodic ($+/-$) boundary conditions
in the compact dimension
\bea
V^+_{\rm eff}(e^{iv})&=&\frac1{L^4}\left[
(2Nn_f\!-\!N^2)\frac{\pi^2}{45}-\frac{Nn_f}{12\pi^2}[v]^2(2\pi\!-\![v])^2\right]\;,\\
V^-_{\rm eff}(e^{iv})&=&V^+_{\rm eff}(-e^{iv})\;,
\eea
where $[v]=v\,\mathrm{mod}\,2\pi$. In the case of periodic boundary conditions the effective potential
has a unique minimum at $[v]=\pi$, while thermal boundary conditions
favor a Polyakov loop on the positive real axis. Gauge invariance then dictates that the actual quantum
vacuum configurations are those center elements of $SU(N)$ closest to $v=0$ and $v=\pi$, respectively.
We thus expect a spontaneous breaking of charge conjugation  for $SU(N)$ with $N=2k+1$, $k\geq1$.

A few comments are in order\footnote{We thank Mithat \"Unsal for pointing those out to us.}:
For gauge group $U(N)$ the phases $v=0$ and $v=\pi$ are elements of the center and thus there
is no spontaneous breaking of charge conjugation for fundamental representation fermions.
This is different from the case of tensor representation fermions, which polarize the Polyakov
line to a phase of $\pm\pi$ and thus always break charge conjugation. Moreover, for large $N$
the vacua for fundamental representation fermions and periodic boundary conditions are separated
from $\pi$ only by $1/N$ effects. Thus the symmetry breaking disappears in the large $N$ limit
together with the difference between $U(N)$ and $SU(N)$ gauge groups and the effects of fundamental
representation fermions in general.

Another calculation, by  Hollowood and  Naqvi \cite{Hollowood:2006cq}, of the phase 
structure of gauge plus 
tensor-representation fermion theories with three small, compact spatial dimensions,
shows the presence of a confinement-deconfinement transition, irrespective of the fermionic
boundary conditions.

The variation on this  explanation which is probably most familiar to a lattice practitioner
comes from strong coupling and large mass expansions.
(A probably incomplete set of early references includes
\cite{Banks:1983me,
Ogilvie:1983ss,
Bartholomew:1983jv,
Green:1983sd,
Hasenfratz:1983ce,
DeGrand:1983fk,
Gocksch:1984yk,
Green:1984ks}.)
Consider a lattice
which is long in the thermal direction and in all but one of
the spatial directions, which we denote by $\hat x$.
There are $N_x$ lattice spacings in the short
direction. We parameterize our fermion action
in a way which is suitable for a hopping parameter expansion,
integrate out the fermions and then expand the determinant
as a power series in the hopping parameter $\kappa$. At large mass,
$\kappa \sim e^{-ma}$ where $a$ is the lattice spacing. In lowest order,
 we find
an effective action whose leading term is proportional to
\be
S_{\rm eff} = \mp \kappa^{N_x} \sum_n {\rm Re} \ \Tr P(\vec n) +\dots
\label{eq:effft}
\ee
where $P(\vec n)$ is the Polyakov line oriented in direction $\hat x$
and  $\vec n$ labels all sites in the lattice with $x=0$.
The sign of the right hand side of Eq.~(\ref{eq:effft}) reflects the
boundary conditions of the fermions in
direction $\hat x$: negative for anti--periodic, positive for periodic.
This result is qualitatively similar to the result of the calculation
of the effective potential: the effect of a short direction
with anti--periodic boundary conditions is to break the center symmetry
to its trivial value, while periodic boundary conditions bias the
Polyakov loop toward a negative value. If $-1$ is an element of the
center, then there is again a unique minimum; otherwise, there will
be degenerate vacua.

We conclude this section with a few simple and general remarks:
\begin{itemize}
\item The actual location of a phase transition (the value of the critical compact radius $L_c$)
can only be determined by a non--perturbative analysis. Its value is a fundamental property
of QCD, like a hadron mass.
\item As the quark mass goes to infinity, $L_c=1/T_c$, the critical temperature for deconfinement
in the quenched theory. 
\item As $m_q\rightarrow 0$ there is presumably a chiral symmetry restoration transition
at some $L_c$. All of the old questions of the interplay of deconfinement and chiral symmetry breaking,
such as whether there are two separate transitions, or one, are present again. \"Unsal and
Yaffe have pictures of proposed phase diagrams with separated transitions.
\item Finally, for SU(3), the pure gauge deconfinement transition is first order. Fermions break the
$Z(3)$ symmetry to $Z(2)$. A first order transition is stable under small symmetry breaking 
effects, so we expect the transition to persist as the quark mass is lowered from infinity.
The transition might end in a second order critical point, or it might
convert to a line of second order points. In either case, we expect that the 
second-order transition would be in
the universality class of the three-dimensional Ising model.
\end{itemize}
\section{Simulations}

\subsection{Methodology}

In our simulations we employ the standard setup for dynamical fermion
lattice simulations; we have three spatial dimensions with periodic boundary
conditions and one time dimension with anti--periodic boundary conditions for the
fermions. The geometry of $\mathbb{R}^3\times\mathcal{S}^1$ is approximated
by keeping all directions but one of the spatial ones (the $\hat x$ direction)
large. We check this by verifying that Polyakov lines wrapping the
large directions are consistent with zero and show no phase preference.
Note that the only difference to a finite temperature simulation is the choice of
boundary conditions in the short direction.

In the present study we choose improved (unrooted, i.e. four--taste) staggered
fermions as a computationally inexpensive way to do simulations with
large dynamical fermion effects. We work with a state-of-the-art improved action to minimize
cutoff effects.
(We have also performed simulations with unimproved staggered fermions. They show similar results
for a transition, but are disfavored for estimating continuum numbers.)
While in principle there are many parameters one
could vary (temperature, length and number
of compact directions, quark mass, flavor number) we will focus on four-flavor
simulations at zero temperature with one compact spatial dimension and only vary its length
and the quark mass. Other choices are briefly discussed in \sect{sec:other}.
We employ the Hybrid Monte Carlo algorithm from the publicly available MILC
code\footnote{\texttt{http://www.physics.utah.edu/\~{}detar/milc/}}
for improved staggered quarks \cite{Orginos:1999cr,Bernard:2001av,Aubin:2004wf}
on a Symanzik gauge background.

\def\arraystretch{1.1}
\TABULAR[t]{|c|l|ccc|ccc|}{\cline{3-8}
\multicolumn{2}{c}{}&
\multicolumn{3}{|c|}{finite $L_x$}&
\multicolumn{3}{|c|}{scale setting}\\\hline
label       &   $am$    &   $N_x$   &   $N_s$   &   $\beta$         &   $N_t$   &   $N_s$   &   $\beta$\\\hline\hline
\textbf{F1} &   0.033   &   6       &   14      &   $6.4\ldots7.4$  &   18      &   10      &   6.0, 6.2, 6.5\\
\textbf{C1} &   0.05    &   4       &   10      &   $5.4\ldots6.4$  &   18      &   10      &   5.7 $\;\ldots\;$ 6.4\\
\textbf{F2} &   0.133   &   6       &   14      &   $5.8\ldots6.7$  &   18      &   10      &   6.4, 6.6, 6.8\\
\textbf{C2} &   0.2     &   4       &   10      &   $5.8\ldots6.9$  &   18      &   10      &   6.1, 6.4, 6.7\\
\textbf{C3} &   0.25    &   4       &   10      &   $5.9\ldots7.0$  &   18      &   10      &   6.2, 6.5, 6.8\\
\textbf{Q}  &   $\infty$&   4       &   8       &   $7.3\ldots8.3$  &   20      &   12      &   7.5, 7.8, 8.1\\
\hline
}
{Simulation parameters.\label{tab:simpar}}
\def\arraystretch{1.0}

The parameters of our simulations are summarized in \tab{tab:simpar}, where the value of
$\beta$ refers to the Symanzik improved gauge action. We ran 500-2000 trajectories
per coupling value for the $N_x=4$ simulations and 200-500 at $N_x=6$. 
In all simulations the tadpole coefficient $u_0$ is fixed to
a value of $0.875$ for all simulations instead of determining it self--consistently at each
value of $\beta$ and $am$. In the scale setting runs we measure HYP smeared 
\cite{Hasenfratz:2001hp} Wilson loops to extract the hadronic scale $r_0$ \cite{Sommer:1993ce}
from the static quark potential \cite{Hasenfratz:2001tw} with results shown in
\fig{fig:scale}. The solid lines are parameterizations of the form
\be
\log(r_0/a)=c_0+c_1\beta+c_2\beta^2\;,
\ee
which we use to interpolate $r_0$. Throughout this work we will assume $r_0=0.5\,$fm.
Note that since we do not tune $u_0$ self--consistently, our bare action differs
from the one used in Ref.~\cite{Gattringer:2001jf} and we therefore obtain rather
different values of $r_0/a$ as a function of $\beta$.

\EPSFIGURE[t]{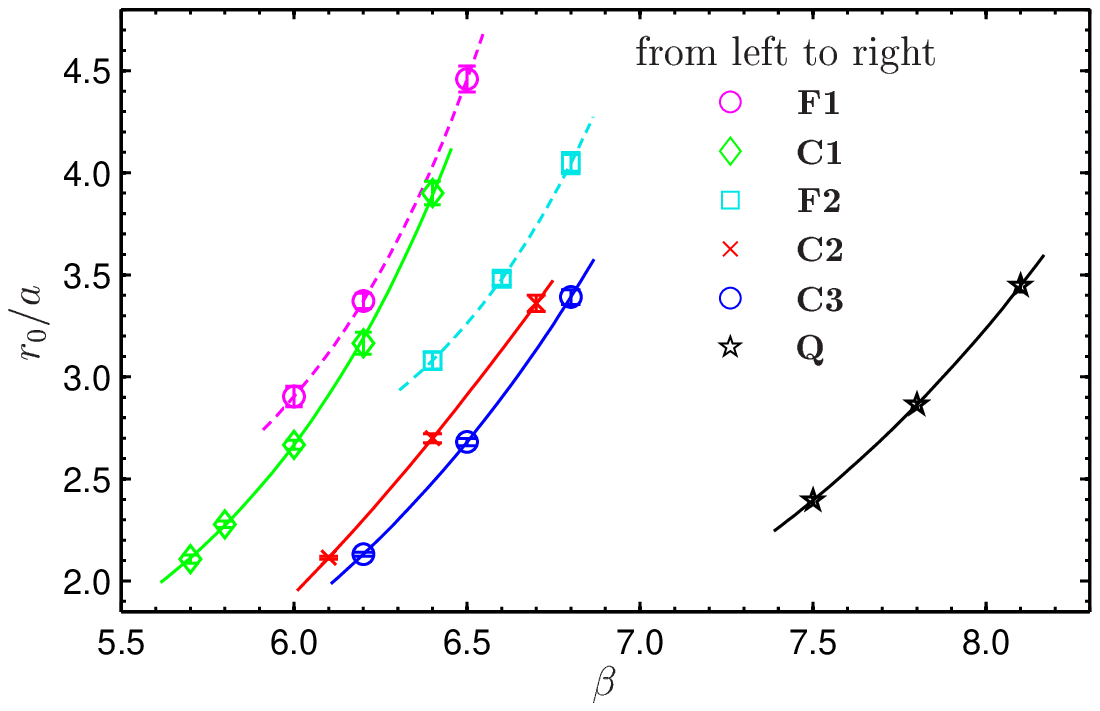, width=.7\textwidth}
{The hadronic scale $r_0/a$ as a function of $\beta$ in the ranges relevant for the
simulations listed in \tab{tab:simpar}.\label{fig:scale}}

\subsection{Results}

We first want to establish that we can indeed find a set of parameters where
the results from \sect{sec:coup} hold qualitatively. To this end we make a comparison: we thermalize
two lattices. One has a compact spatial direction, and our initial gauge configuration
was fixed to the identity. The second configuration has a compact temporal direction, and its
initial gauge fields had a time--like Polyakov loop with a phase of $\exp(-2\pi i/3)$.
We choose $N_{t/x}\!=\!4$ for the compact direction, which corresponds to $\simeq0.6\,$fm at $\beta=6.65$
and a quark mass of $am\!=\!0.2$. In both cases the non--compact directions all have 10 lattice sites.

In \fig{fig:therm} the direction of thermalization is indicated by an arrow
for both cases. One can clearly see that the finite temperature simulation
(left panel) finds a stable minimum on the real axis. The accumulation
on the complex $Z(3)$ leaf (indicated by a circle) might indicate a meta--stable
state. The simulation with compact \emph{spatial} dimension (right panel) shows the
opposite behavior. After following the real axis to the origin it settles
to an equilibrium (indicated by a dot) with a phase around $-2\pi/3$.
Thus we encounter precisely the behavior that the analytical arguments from
\sect{sec:coup} suggest and now proceed to a more detailed study of the transition.

\EPSFIGURE[t]{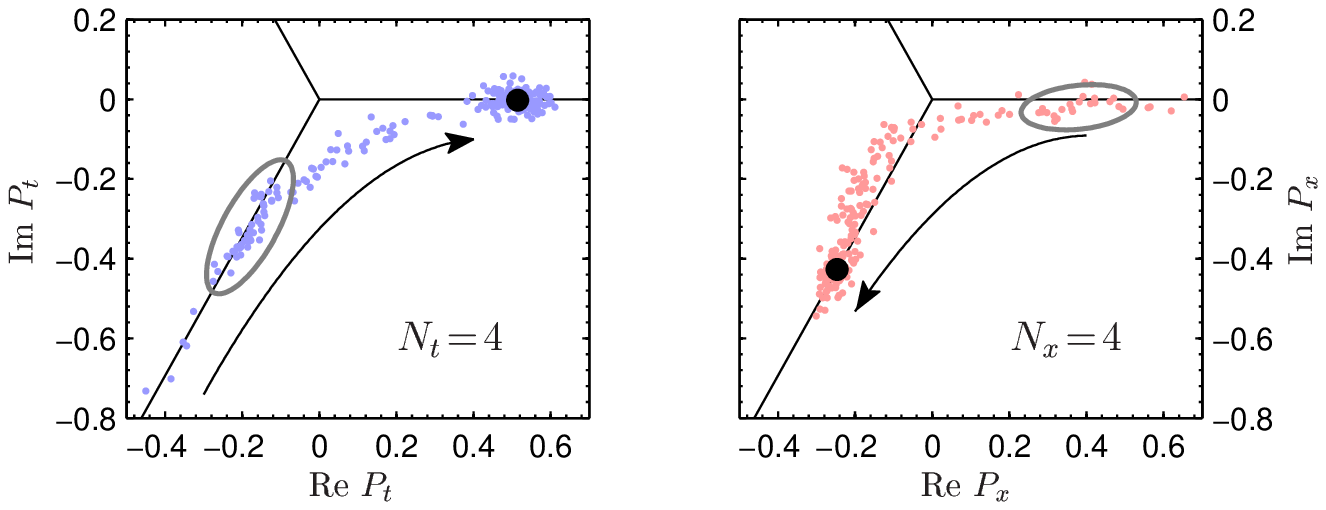, width=.9\textwidth}
{Thermalization histories of the Polyakov line in the short direction. Left panel:
short time direction. Right panel: short spatial ($x$) direction.\label{fig:therm}}

We focus on simulations with one compact spatial dimension ($x$) at zero temperature
and vary the extent of the compact dimension by changing $\beta$. Fig.~\ref{fig:scat} shows
scatter plots of $P_x$ after equilibration from sets \textbf{C2}. For better readability each point represents
the average of 5 consecutive measurements of $P_x$. The length of the compact dimension
is indicated in units of $r_0$. The numerical data confirms the expectation that
the fermion determinant pushes the Polyakov lines toward the negative real axis.
At the largest extent we simulated with $N_x\!=\!4$, corresponding to $L_x=2.48\,r_0$, $P_x$
is centered at -0.0275(17) on the real axis. With decreasing length, it moves to more negative
values and spreads in the imaginary direction toward the complex center elements.

Below a certain critical length, the distribution becomes disjoint and one of the complex
$Z(3)$ elements is selected as the ground state configuration with only the magnitude
of $P_x$ increasing with decreasing $L_x$. Thus, for $SU(3)$ the fermion polarization
spontaneously breaks charge conjugation invariance, indicated by a non-vanishing imaginary part of the
Polyakov loop. With the chosen algorithm and at the volumes we simulated, no tunneling events
between the two vacua were observed. This situation changes in finite temperature simulations
and/or with more than one compact dimension, see \sect{sec:other}.

\DOUBLEFIGURE[t]
{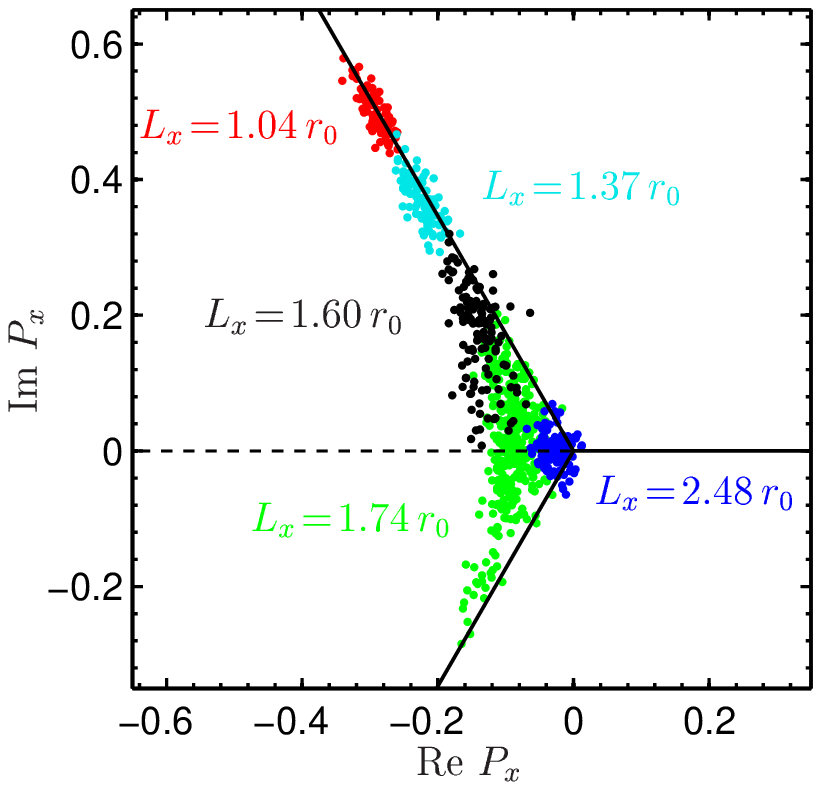, width=.48\textwidth}
{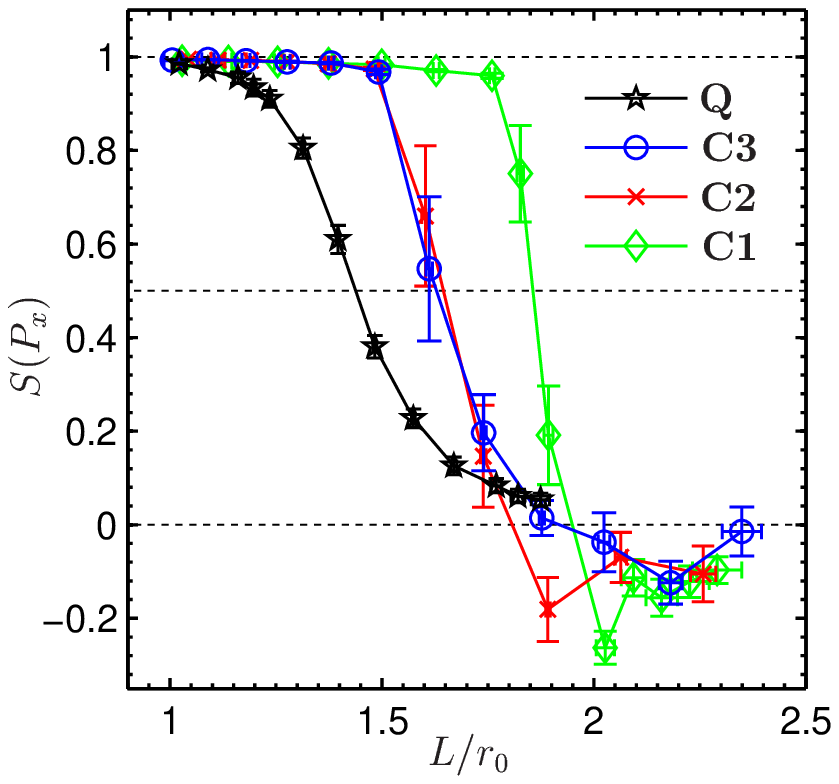, width=.48\textwidth}
{Equilibrium distributions of the Polyakov line from \textbf{C2} for varying extent
of the compact spatial dimension.\label{fig:scat}}
{The observable $S$ from Eq.~(\protect\ref{defS}) for the $N_x\!=\!4$ simulations.\label{fig:Sone}}

In principle the imaginary part of the Polyakov loop
could be used to make quantitative statements about the transition. However,
close to the critical length the increased spread of the Polyakov lines (see e.g. the $L_x\!=\!1.74$
data in \fig{fig:scat}) would only affect the variance of the imaginary part.
An observable better suited to make quantitative statements is the correlation
of the phase of the Polyakov loop with that of one of the group centers $e^{2k\pi i/3}$. 
Another advantage is that, unlike the imaginary part, the phase of the Polyakov line does
not require renormalization.
We map the phase range between two $SU(3)$ center elements to the full circle by
taking $(P/|P|)^3$ and then project onto the real axis,
\be
S(P)=\re (P/|P|)^3=\cos(3\arg P)\;. \label{defS}
\ee
The expectation value of $S(P)$ is close to one in the broken phase, when all Polyakov lines are in the
vicinity of one of the $Z(3)$ centers. It is zero for randomly distributed phases and negative if the phases
lie between the $Z(3)$ elements, e.g. around the negative real axis.
(This observable is reminiscent of a similar one used to locate the pure gauge transition
\cite{Christ:1985wx}.)   
The center correlation $S(P_x)$ for the $N_x\!=\!4$ simulations is shown in \fig{fig:Sone}
as a function of the length of the compact dimension. For the dynamical simulations
we observe that the value of $S$ abruptly decreases from almost unity to small negative values before
approaching zero in the large volume limit. For the heavier data sets \textbf{C3} and \textbf{C2},
the transition occurs at $\simeq0.8\,$fm and for the lightest set \textbf{C1} with $am\!=\!0.05$
the critical length is slightly larger at $0.93\,$ fm. The extraction of the critical length
is discussed in the next section.

For comparison we also show the quenched results, run \textbf{Q}. As discussed above, the quenched
limit of our setup is just the familiar finite temperature phase transition for the pure gauge theory.
Here the behavior is qualitatively different as all \emph{three} center elements are equivalent vacua in the
broken phase. Moreover, the heatbath/overrelaxation algorithm allows tunneling between those.
With our chosen observable the transition appears much smoother than in the presence of dynamical fermions.

To establish that we are indeed seeing
a \emph{physical} effect and not just lattice artifacts, we simulated finer lattices ($N_x\!=\!6$).
 Data from the $N_x\!=\!6$ simulations are shown in Figs.~\ref{fig:scatfine} and
\ref{fig:Stwo}.

\DOUBLEFIGURE[t]
{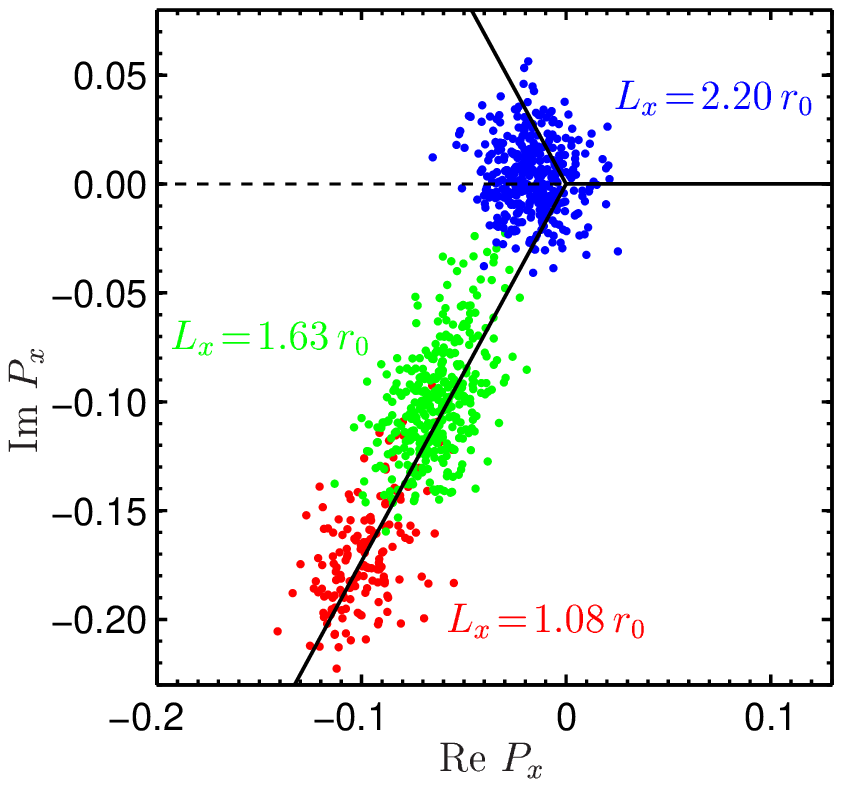, width=.48\textwidth}
{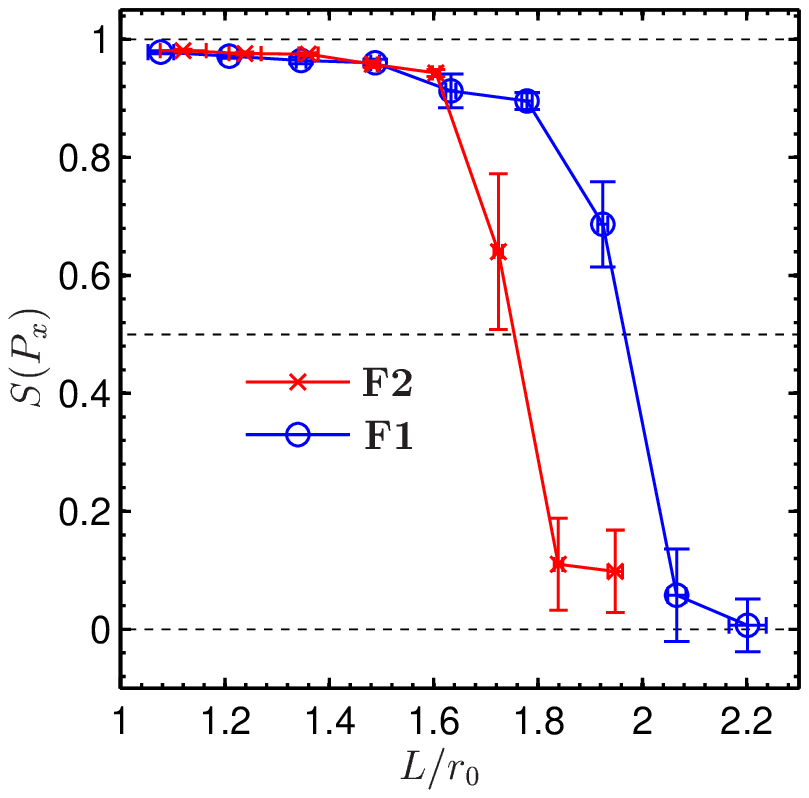, width=.48\textwidth}
{Equilibrium distributions of the Polyakov line from \textbf{F2}.\label{fig:scatfine}}
{The center correlation $S$ from Eq.~(\protect\ref{defS}) for the $N_x\!=\!6$ simulations.\label{fig:Stwo}}

The results from the finer lattice spacing are qualitatively similar to the $N_x\!=\!4$ data
presented in Figs.~\ref{fig:scat} and \ref{fig:Sone}. The transition is still present and only
slightly more rounded. The critical length has moved to somewhat larger
values and there is no longer a region of negative values of $S(P_x)$ just above the transition.
One might thus interpret this behavior as a lattice artifact that disappears when going from
$N_x\!=\!4$ to 6.

\EPSFIGURE[t]{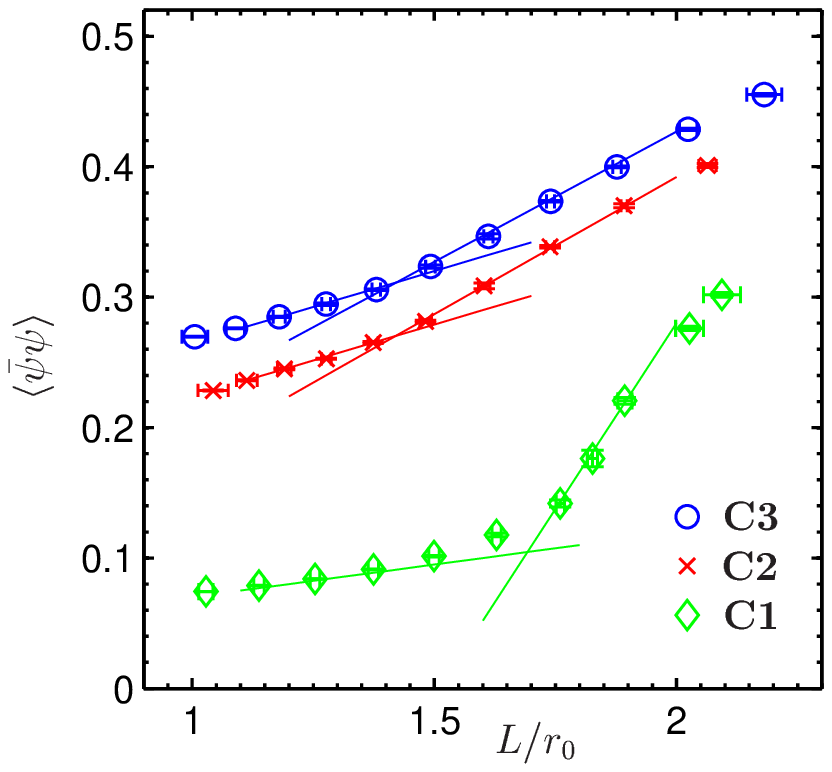, width=.4\textwidth}
{The scalar condensate $\langle\bar\psi\psi\rangle$ vs. the extent of the compact dimension
for the $N_x\!=\!4$ runs.\label{fig:pbp}}

Finally, we consider the chiral condensate $\langle\bar\psi\psi\rangle$.
A non--vanishing expectation value indicates spontaneous breaking of chiral symmetry,
which occurs in large volumes.
Finite volume, non--zero quark mass as well as discretization errors tend to wash out this transition
such that only a smooth crossover is observed.
As we are only interested in the qualitative behavior across the transition, we consider the
unrenormalized lattice condensate. The results from the $N_x\!=\!4$ runs, again as a function of
$L/r_0$, are shown in \fig{fig:pbp}. As expected, for the heavier masses (\textbf{C2}, \textbf{C3})
no strong signal is observed. However, as marked by the lines, there is an inflection point, where
the slope (as a function of $L_x$) increases as the compact dimension gets bigger.
The former is located just below $L\!=\!1.5\,r_0$, which
is where the center correlation (see \fig{fig:Sone}) starts to drop. The lightest mass (\textbf{C1})
shows a much stronger crossover and again the inflection point agrees with the data from the center
correlation. Almost identical results are obtained at the finer lattices.

Thus the data indicate that in $N_f=4$ QCD, at zero temperature 
with one compact spatial dimension, there is
a single crossover, i.e. a critical length, below which chiral symmetry is restored and charge
conjugation is spontaneously broken.

\section{Interpretation}

As a definition of the critical coupling $\beta_c$  we take the value, where the center correlation
$S(P)\!=\!0.5$. In practice we linearly interpolate between the two values of $\beta$
above and below this point. Since this quantity shows a sharp transition the results will not change
significantly when different definitions are adopted. The values of $\beta_c$ and the critical length
$L_c$ for all runs are given in \tab{tab:crit}. The error on $L_c$ includes the (statistically independent)
error of $r_0$.

\TABULAR[t]{|c|l|clll|}{
\hline
label       &   $am$    &   $N_x$& $\quad\ \beta_c$   &   $\ L_c/\,$fm & $L_c^{-1}/\,$MeV   \\\hline\hline
\textbf{F1} &   0.033   &   6    & 6.070(12)   &   0.983(14)         & $\ $ 201(9)  \\
\textbf{C1} &   0.05    &   4    & 5.728(9)    &   0.928(9)          & $\ $ 213(4)  \\
\textbf{F2} &   0.133   &   6    & 6.574(20)   &   0.877(14)         & $\ $ 225(12) \\
\textbf{C2} &   0.2     &   4    & 6.269(25)   &   0.822(17)         & $\ $ 240(12) \\
\textbf{C3} &   0.25    &   4    & 6.387(34)   &   0.814(22)         & $\ $ 242(16) \\
\textbf{Q}  &   $\infty$&   4    & 7.752(11)   &   0.719(6)          & $\ $ 274(6)  \\
\hline
}
{Results for the critical length.\label{tab:crit}}
\def\arraystretch{1.0}

Taking into account the rather coarse lattice spacing and the arbitrariness in the
precise definition of the critical length, the result for the quenched case
is in good agreement with other determinations of the
quenched critical temperature (see e.g. 
\cite{Karsch:1999vy}, who quotes $T_c \sim 270$ MeV.)
Comparing the data
from the coarse dynamical lattices in \tab{tab:crit} we see that the critical length
increases with decreasing quark mass. Data at fixed $N_x\cdot am=mL_c$, i.e. \textbf{C1}/\textbf{F1}
and \textbf{C2}/\textbf{F2}, suggest that scaling violations are not large and that
the lattice artifacts decrease the value of $L_c$. An extrapolation assuming $a^2$ scaling gives
continuum values of $L_c\!=\!1.027(26)\,$fm and $0.922(29)\,$fm for $mL_c=0.2$ and 0.8, respectively.

Using all the $N_x\!=\!4$ data, we obtain a phase diagram of the critical length as a function of
the quark mass as shown in \fig{fig:phasediag}. The critical length increases as the quark mass
is lowered from infinity and although we have no data at very light quark masses, the dependence
does not seem to be very strong. Clearly, more quantitative statements will require simulations
at finer lattice spacing and more (and lighter) quark masses.

\DOUBLEFIGURE[t]
{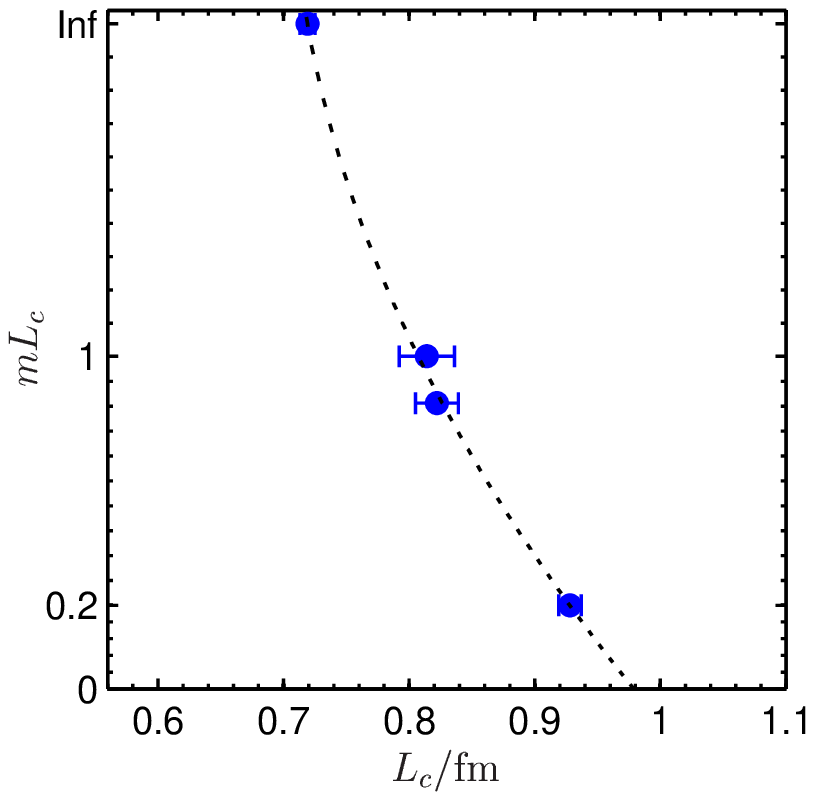, width=.44\textwidth}
{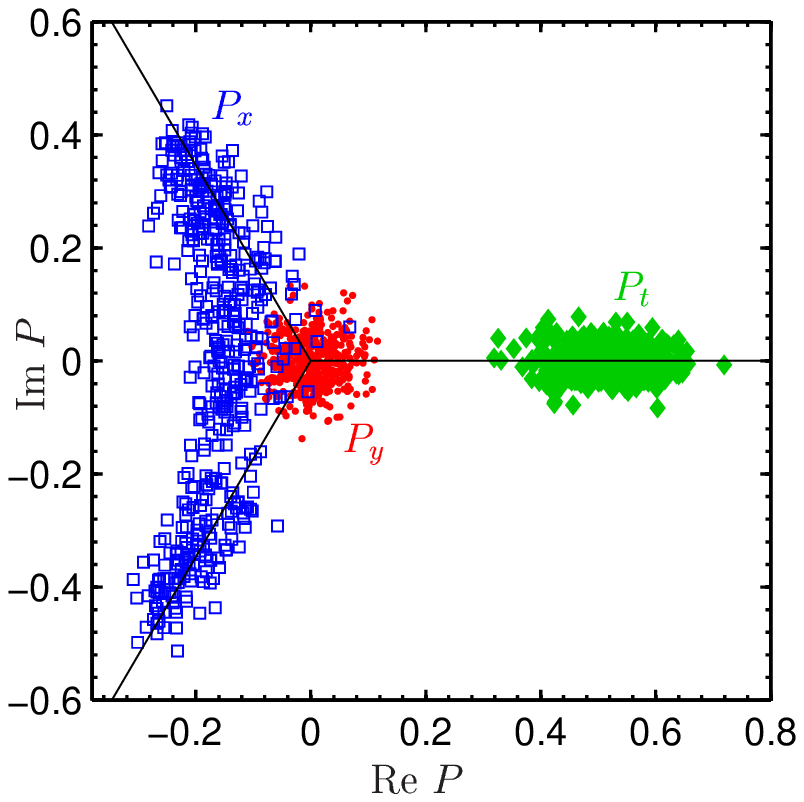, width=.44\textwidth}
{The phase diagram at $N_x\!=\!4$, where we have three dynamical and a quenched data point
for the critical length of the compact dimension.\label{fig:phasediag}}
{Equilibrated Polyakov lines for $\beta\!=\!6.65$, $am\!=\!0.2$ and $[N_x,N_y,N_z,N_t]\!=\![4,10,10,4]$.
\label{fig:hotsmall}}

\subsection{Other geometries}
\label{sec:other}

To briefly explore other geometries, we chose a parameter set in the broken phase ($am\!=\!0.2$,
$\beta\!=\!6.65$) and simulated all possible combinations of compact ($N\!=\!4$) and non--compact
($N\!=\!10$) dimensions, including finite temperature.
In all cases we confirm our expectation that compact spatial dimensions support a Polyakov loop
in one of the two complex $Z(3)$ directions while a compact time direction 
produces a positive Polyakov loop.
If more than one spatial dimension is short, the bulk volume is reduced and we observe tunneling between
the two states. The values of the Polyakov loops in the compact direction are not correlated,
showing that indeed all vacua are equivalent.

At finite temperature already one spatial dimension allows tunneling of the Polyakov line in the compact
space dimension. \fig{fig:hotsmall} shows Polyakov lines from such a geometry. While those from the
non--compact dimensions cluster around zero ($P_y$; $P_z$ not shown) the time--like loops have a positive
mean value and those in the compact spatial dimensions spread along the complex $Z(3)$ directions. It is not
known whether the transition persists at high temperature and whether the finite temperature transition at
$1/L_x\!=\!0$ and the $C$ breaking transition at zero temperature are connected. However, an exploration
of the phase diagram of inverse temperature and compact length might be interesting.

\section{Summary}

For the usual kind of QCD simulations, where the goal is to model QCD in infinite volume,
the $C$-breaking transition is just an artifact which needs to be checked. For example,
 \"Unsal and Yaffe commented that our study of the $N_f\!=\!1$ condensate might have been done
in the $C$-broken phase. We checked our data sets for our studies of the $N_f\!=\!1$ and 2
condensates \cite{DeGrand:2006uy,DeGrand:2006nv} and verified that we were in the $C$-symmetric phase.

In the present paper we studied QCD with dynamical quarks with one compact spatial dimension.
We find a fundamental length scale, the critical length $L_c$, below which QCD with four
dynamical quark flavors shows spontaneous breaking of charge conjugation.
The continuum value of $L_c$ is approximately 1 fm. It seems to be connected to the finite
temperature transition in the limit of infinite quark and persists down to the smallest
quark masses we simulated. There we also observe a crossover in the chiral condensate, signaling
the restoration at chiral symmetry at the same critical length. We have not determined 
the order of the transition: this would require more extensive finite-size scaling studies.

The spontaneous breaking of charge conjugation is expected from both perturbative arguments
and a strong coupling expansion. For technical reasons we performed simulations with four quark
flavors but do not expect the conclusions to change significantly for other choices.
(We have done some pilot studies with $N_f=2$ flavors of improved Wilson fermions,
and have also seen the crossover to a C-broken phase.) Clearly we are seeing only the tip of
 the iceberg: just as for the finite-temperature transition, the transition is affected
by the number of colors and flavors, group representation of the fermions, and the relative
sizes of all small compact dimensions.

\acknowledgments
We are grateful to Mithat \"Unsal and  Larry Yaffe 
for correspondence which initiated this investigation,
and to Mithat \"Unsal for a critical reading of the manuscript.
Some of the simulations were performed on the cluster at Fermilab.
This work was supported in part by the US Department of Energy.

\appendix

\bibliography{/axp/huron/wrk1/ltm/hoffmann/tex/refs}           
\bibliographystyle{JHEP}

\end{document}